\documentclass{vgtc} 

\PassOptionsToPackage{svgnames}{xcolor}
\usepackage[svgnames,dvipsnames]{xcolor}

\graphicspath{{figs/}{figures/}{pictures/}{images/}{./}} 

\usepackage{times}                     

\usepackage{tabu}                      
\usepackage{booktabs}                  
\usepackage{lipsum}                    
\usepackage{mwe}                       

\usepackage{booktabs}
\usepackage{multirow}

\usepackage{graphicx}
\usepackage{tikz}
\usepackage{xcolor}

\usepackage{mathptmx}                  

\onlineid{1218}

\vgtccategory{Empirical Study}

\vgtcinsertpkg

\title{Capturing Visualization Design Rationale}

\author{
\makebox[\textwidth][c]{%
\begin{minipage}{1.2in}\centering
Maeve Hutchinson\thanks{maeve.hutchinson@citystgeorges.ac.uk}\\
\scriptsize giCentre,\\City St George's,\\University of London
\end{minipage}\hfill
\begin{minipage}{1.2in}\centering
Radu Jianu\\
\scriptsize giCentre,\\City St George's,\\University of London
\end{minipage}\hfill
\begin{minipage}{1.2in}\centering
Aidan Slingsby\\
\scriptsize giCentre,\\City St George's,\\University of London
\end{minipage}\hfill
\begin{minipage}{1.2in}\centering
Jo Wood\\
\scriptsize giCentre,\\City St George's,\\University of London
\end{minipage}\hfill
\begin{minipage}{1.5in}\centering
Pranava Madhyastha\thanks{pranava.madhyastha@citystgeorges.ac.uk}\\
\scriptsize City St George's,\\University of London;\\The Alan Turing Institute
\end{minipage}
}
}

\teaser{
  \centering
  \includegraphics[width=\linewidth]{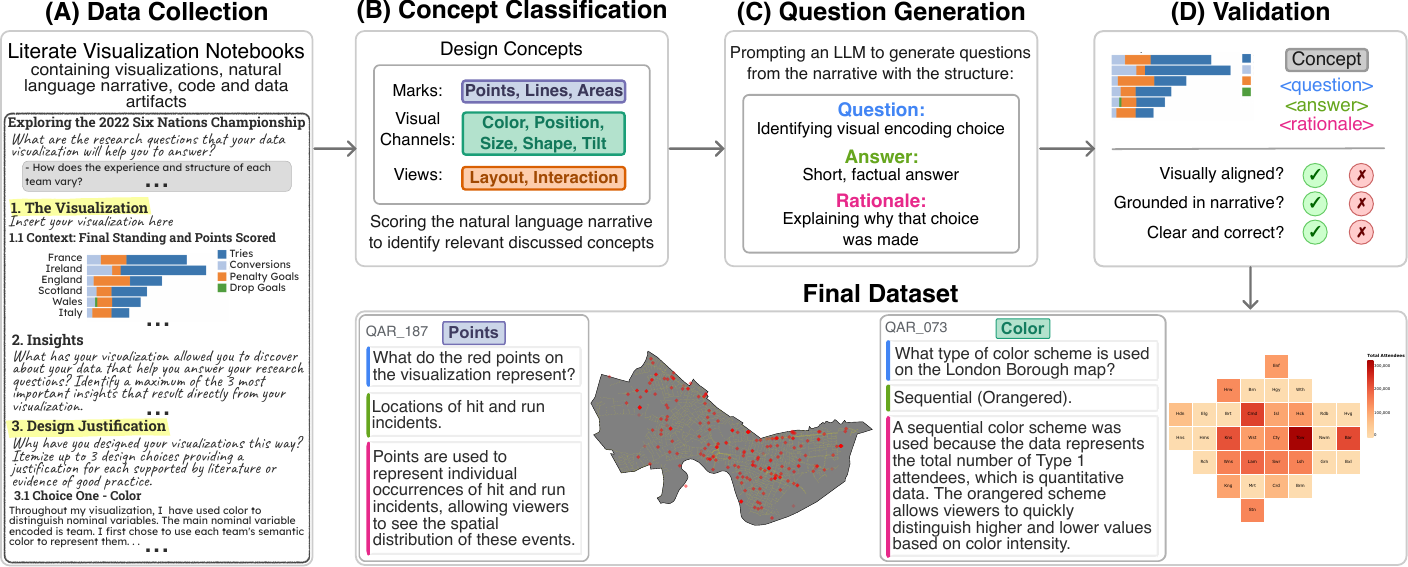}
  \caption{Overview of the structure of our study, showing (A) an example of a student-authored literate visualization notebook, and (B) the ten visualization design concepts used to classify rationale. Together, these components frame our methodology for (C) extracting and (D) validating question, answer, and rationale (QAR) triples from real-world student visualization design narratives.}
  \label{fig:teaser}
}

\abstract{
Prior natural language datasets for data visualization have focused on tasks such as visualization literacy assessment, insight generation, and visualization generation from natural language instructions. These studies often rely on controlled setups with purpose-built visualizations and artificially constructed questions. As a result, they tend to prioritize the interpretation of visualizations, focusing on decoding visualizations rather than understanding their encoding. In this paper, we present a new dataset and methodology for probing visualization design rationale through natural language. We leverage a unique source of real-world visualizations and natural language narratives: literate visualization notebooks created by students as part of a data visualization course. These notebooks combine visual artifacts with design exposition, in which students make explicit the rationale behind their design decisions. We also use large language models (LLMs) to generate and categorize question-answer-rationale triples from the narratives and articulations in the notebooks. We then carefully validate the triples and curate a dataset that captures and distills the visualization design choices and corresponding rationales of the students.
} 


\begin{document}


\firstsection{Introduction}

\maketitle


A growing body of visualization research investigates the multifaceted roles of language --- from its use within charts as annotations and titles \cite{stokes_striking_2022}, to enabling user interaction via queries or commands \cite{dibia_lida_2023}, and facilitating communication about visualizations, including articulating interpretations or design choices. This exploration has resulted in the development of a variety of datasets capturing these diverse language related aspects of visualization practice and understanding. 
Existing research, for instance, has examined visualization literacy --- the ability to read and interpret information presented 
graphically \cite{lee_vlat_2017}. Datasets in this space, such as VLAT \cite{lee_vlat_2017}, often 
assess \textit{visualization} literacy through questions 
that probe how effectively users or models can extract information from visualizations. This focus centers primarily on \textit{decoding} visual representations.

Similarly, Chart Question Answering (CQA), a related object of study in natural language processing, develops 
datasets comprising 
questions about the information conveyed in 
visualizations \cite{kim_answering_2020,kafle_dvqa_2018,methani_plotqa_2020}. 
Like literacy assessment, CQA predominantly emphasizes the decoding aspect of visualizations, evaluating comprehension rather than the underlying design principles.
A common characteristic of many existing datasets is their construction within controlled settings. Often designed to isolate specific phenomena, they may employ purpose-built visualizations, rely on crowdsourced responses to templat-based prompts \cite{srinivasan_collecting_2021}, or utilize synthetically generated queries \cite{lee_vlat_2017}. Consequently, resources capturing the nuances of language used in more authentic, ecologically valid visualization practices remain relatively scarce.

In this paper, we focus on capturing the \textit{human reasoning} underlying the visualization design process itself, where we exploit the textual articulations, justifications, and narratives provided by designers. We introduce a dataset specifically curated to capture these design rationales \textbf{in the wild}, derived from authentic visualization activities conducted by students learning visualization principles. This approach aims to surface the genuine considerations that inform encoding choices, offering a perspective grounded in practice rather than in controlled or synthetic environments.
While there has been some work that has aimed to capture encoding principles, notably Draco \cite{moritz_formalizing_2019}, which formalized visualization design knowledge as a structured set of constraints compiled from theory. While valuable, this approach represents best practices in a rule-based format. We believe that our work complements this line of work by capturing encoding rationale as expressed through language in real-world scenarios, preserving the potential ambiguities and situated reasoning inherent when applying design principles in practice.

To access expressions of this reasoning, we draw from Literate visualization (litvis)\cite{wood_design_2019}. litvis promotes the integration of code, visualizations, and textual explanation within a single document (a `litvis notebook'). This format inherently encourages designers to articulate their rationale alongside the construction of visualizations. The litvis notebooks used in our study followed a narrative schema specifically prompting students to justify their design decisions. In the following sections, we detail how we collected and structured data from these notebooks to create a novel dataset. 

\section{Methodology}
\subsection{Data Collection}

Our dataset originates from litvis notebooks created by both undergraduate and postgraduate students as a part of their final coursework for a 10-week long data visualization course at our university. The course covered fundamental principles of data visualization design and their implementations. 
Our study and the data collection process received formal approval from our university's Research Ethics Committee. Following the approval, graduated students were informed about our study and their explicit informed consent was sought for the use of their coursework. Our dataset is fully sourced from the submissions of students who duly provided permission for their materials to be processed for this research.

The assessed coursework required students to select a dataset, formulate research questions, and design custom visualizations intended to answer those questions. These were submitted as litvis notebooks: markdown documents integrating textual narrative, analysis datasets, code blocks in Elm, and inline visualizations rendered via \texttt{elm-vegaLite} \cite{noauthor_gicentreelm-vegalite_2024} (an example notebook is shown in Figure \ref{fig:teaser} A). Crucially, these notebooks followed a narrative schema designed for the course, which included a `Design Justification' component. This section explicitly asked students: “Why have you designed your visualizations this way?” and prompted them to “Itemize up to 3 design choices providing a justification for each supported by literature or evidence of good practice.” This literate visualization environment, combined with the specific instructions, encouraged students to surface design rationale choices that might otherwise remain implicit.

We excluded submissions unsuitable for our analysis, specifically those lacking a successfully rendered visualization, containing personally identifiable information, offering insufficient textual justification regarding design decisions, or otherwise failing to meet a minimum quality threshold. After this initial filtering, 22 notebooks were retained for further processing.

From each suitable notebook, we extracted two primary types of content: the textual justifications and the corresponding visualizations. All language associated with the `Design Justification' field was programmatically extracted from the markdown source. Visualizations were captured from the rendered HTML version of the notebooks using a headless browser.
We divided each student's justification text into segments of up to 200 words to prepare it for subsequent processing by language models.

\subsection{Concept Classification}

To systematically analyze the design justifications and enable targeted question generation, we defined a focused set of core design concepts. These concepts, informed by visualization theory, provide a structured lens through which to interpret the articulated rationale within the collected text segments.




Our conceptual framework builds upon foundational visualization principles, drawing from Bertin's original concepts of graphical elements and retinal variables \cite{bertin_semiology_2011} and their subsequent refinement into mark types and visual channels by Munzner \cite{munzner_visualization_2014}. We pragmatically aligned and refined Munzner's conceptualization based on the specific mark types and encoding channels supported by the Vega-Lite grammar \cite{satyanarayan_vega-lite_2017}, as students implemented their visualizations using \texttt{elm-vegaLite}. Beyond basic encoding elements, we also incorporated concepts related to the higher-level composition of visualizations (Layout) and user interaction capabilities (Interaction), as these were prominent themes in the student work and course curriculum. This process, which balances theoretical grounding with the practicalities of the students' implementation environment, yielded 10 core design concepts organized into three broad categories (summarized in Figure \ref{fig:teaser}B). 
%


We note that this is neither a design space nor a comprehensive taxonomy of visualization concepts. Rather, it is tailored to the expressivity of \texttt{elm-vegaLite}, in particular the kind of visualizations and concepts that the students studied. 
%
%
%
These concepts are not mutually exclusive. Marks necessarily encode data through visual channels. However, the rationale for choosing a certain mark type may differ from the rationale for encoding data through a certain visual channel. Our conceptualization is intended to capture these levels of design decision-making, from low-level components (marks) to the high-level composition of visualizations (views).

To systematically associate the extracted text segments with these design concepts, we employed LLooM \cite{lam_concept_2024}, a Large Language Model (LLM)-based algorithm developed for concept induction from unstructured text. For each of our 10 design concepts, we crafted a descriptive prompt defining the concept and outlining relevance criteria. LLooM's scoring function took each text segment and evaluated it against each concept prompt, utilizing an LLM to assign a relevance score between 0 (not relevant) and 1 (highly relevant). We adopted a strict threshold: only concepts achieving the maximum score of 1 for a given text segment were associated with that segment. As student justifications often covered multiple design aspects simultaneously, a single text segment could be tagged with multiple concepts under this approach. 

This concept scoring process served two crucial purposes for our study: a) It identified the specific concepts discussed in each text segment so we could direct the question generation phase (discussed in \cref{sec:qgen}), minimizing the generation of questions unrelated to visualization design; and b) it allowed us to analyze the distribution and nature of design rationale across these different concepts (we present this in \cref{sec:dataset}).

\subsection{Question Generation} \label{sec:qgen}


We construct the dataset as triples consisting of a question, an answer, and a rationale (QAR). This aligns with established use of question answering formats in datasets linking language and visualization \cite{lee_vlat_2017, kafle_dvqa_2018, bendeck_empirical_2024}. Our dataset design is further informed by the two-stage structure employed in the Visual Commonsense Reasoning dataset \cite{zellers_recognition_2019}, which first asks a question identifying elements in a picture, and then asks for a rationale behind the answer.  Similarly, each of our questions is framed as a simple identification query about what aspect of visual encoding or design is being discussed, followed by a rationale that explains why that particular encoding decision was made. Thus, each triple targets two aspects of visualization design: the identification of a design choice and a justification for that choice. This structure enables us to systematically surface the underlying design rationales embedded within students’ natural language narratives

Our question generation process relies exclusively on the natural language text, without using the visualizations themselves. This approach is inspired by Changpinyo et al. \cite{changpinyo_all_2022}, who demonstrated a method for constructing visual question answering (VQA) datasets automatically and at scale from only image captions, without access to the images themselves. Analogously, we leverage students’ detailed descriptions of their visualizations and design to generate questions that surface their design rationale.

To generate QAR triples we prompted an LLM with the text segments and the relevant design concepts --- those assigned a relevance score of 1 from LLooM \cite{lam_concept_2024}. Each prompt also included a brief description of the concept, example questions to guide generation, and instructions to the model. The model was instructed to generate two QAR triples per concept per text segment, drawing exclusively on the student text without incorporating external information. Through this process, we generated 362 QAR triples which then underwent a stringent procedure for manual validation.

\subsection{Human Validation}

To ensure the quality and reliability of the dataset all 362 LLM-generated QAR triples underwent rigorous human validation.
Each triple was evaluated against predefined rejection criteria, designed to address potential sources of error introduced during data collection, concept scoring, or QAR generation. These criteria fell into three main categories: (1) misalignment with available visualizations, (2) quality issues originating from the LLM generation process, and (3) quality issues stemming from the original student text. A key part of this validation process also involved associating each valid QAR triple with the specific visualization(s) it described, as students often produced multiple visualizations within a single notebook.

The first criterion for rejection concerned the alignment between the QAR triples and the available visualizations. While students learned to use \texttt{elm-vegaLite} during the course, its use was not mandatory in the final submission; sketches and mock-ups were permitted and often encouraged in cases where students were unable to encode their intended designs. These sketches fall outside the scope of our dataset. Additionally, in some cases, visualizations failed to render due to the unavailability of the underlying datasets. Thus, since we generated the triples from the text alone, several triples referred to visualizations that we did not recover through our data collection process. If a triple was not related to any of the available visualizations, it was excluded. Visual alignment issues accounted for 44 (31.2\%) of rejections.

The second rejection criterion concerned data quality issues introduced by LLMs, mostly during the question generation process. Generally this involved the model also introducing information in the generated triple not present in the original text. Sometimes this information was hallucinated entirely, producing a rationale that was not present in the text and did not align with visualization theory. Other times, the model generated fabricated rationales that were plausible, reflecting sound visualization principles but were not grounded in the text. These triples were rejected as we are aiming to surface real rationales produced by the students. 
Finally, in rare instances, concept misclassifications at the scoring stage led to attempted generation about a concept not discussed in the text segment. Such cases were rare and the LLooM \cite{lam_concept_2024} scoring function generally performed well. Model-related issues accounted for 73 (51.7\%) of rejections.

The final rejection criterion concerned quality issues in the original text. Although the students had been trained in visualization, they are not experts, and so variability in the student natural language occasionally led to quality issues in the generated QAR triples. In some cases, students mischaracterized their visualizations, using either inaccurate language to describe their design choices. In other cases, imprecise or ambiguous language resulted in answers or rationales that are too vague to provide useful insights. 
It is important to note that not all visualizations accepted into the final dataset represent ``theory-perfect'' samples. The dataset reflects the authentic, and sometimes imperfect, design decisions of students. Text-related issues accounted for 24 (17.0\%) of rejections. This comparatively low proportion likely reflects our initial quality filtering.


Following our comprehensive human validation process, we retained 221 high-quality QAR triples, corresponding to 124 visualizations. This represents a final acceptance rate of 61.0\% from the initial LLM-generated pool, forming the curated dataset analyzed in the subsequent sections.

\section{Dataset} \label{sec:dataset}
We now present an analysis of the curated dataset by examining the surfaced answers and rationales across the defined visualization concepts. Table \ref{tab:qar-distribution} provides an overview of the composition of the final dataset. 

\begin{table}[h]
\caption{Distribution of accepted QAR triples across visualization concepts.}
\centering
\begin{tabular}{llrr}
\toprule
\textbf{Category} & \textbf{Concept} & \textbf{Count} & \textbf{\% of Total} \\
\midrule
\multirow{3}{*}{Marks} 
  & Areas  & 28 & 12.7\% \\
  & Points & 26 & 11.7\% \\
  & Lines  & 4  & 1.8\% \\
\midrule
\multirow{5}{*}{Visual Channels} 
  & Color    & 53 & 24.0\% \\
  & Position & 15 & 6.8\% \\
  & Size     & 12 & 5.4\% \\
  & Shape    & 6 & 2.7\% \\
  & Tilt     & 0  & 0.0\% \\
\midrule
\multirow{2}{*}{Views} 
  & Interaction & 39 & 17.6\% \\
  & Layout      & 38 & 17.1\% \\
\midrule
\textbf{Total} &  & \textbf{221} & \textbf{100.0\%} \\
\bottomrule
\end{tabular}
\label{tab:qar-distribution}
\end{table}

\begin{figure}[ht!]
\includegraphics[width=1\columnwidth]{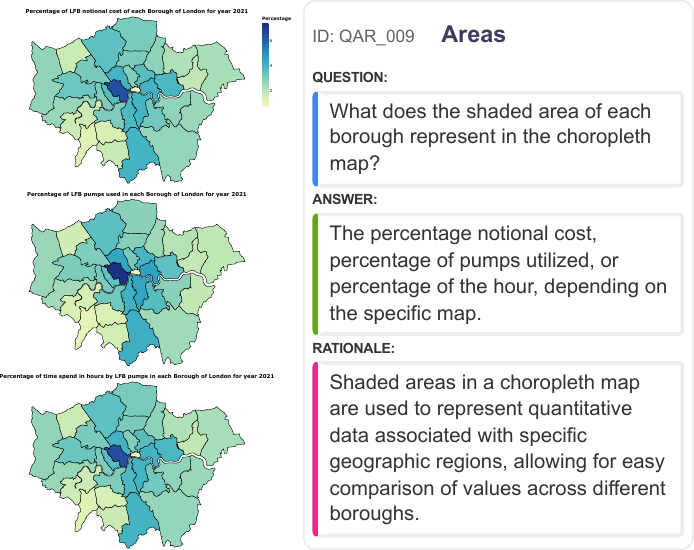}
  \caption{Example of an `Areas' Question, Answer and Rationale from our dataset.}
  \label{fig:qar_area1}
\end{figure}

\noindent{\textbf{Areas}}. This category predominantly features choropleth maps and other geospatial visualizations using area representations. A number of students justify the use of colored areas in this way, as in \cref{fig:qar_area1}.
Several triples surface discussions of distorted or relaxed geographies, where students considered whether conventional map boundaries or abstracted regions better supported their goals. 
Beyond geospatial visualizations, several triples identify the use of area to encode data, for example, through the use of bubble plots or Sankey diagrams. Some of these triples identified perceptual issues with using area to encode data, and explored the use of non-linear scaling to mitigate these issues.


\noindent{\textbf{Points}}. Triples about points most frequently discuss scatterplots to visualize the relationship between two quantitative variables, with questions and answers typically identifying what the data points represent. The rationales often highlight the suitability of scatterplots for identifying trends, clusters, or outliers. Several triples also identify the spatial application of point marks, for example, to show the locations of schools or hit-and-runs on a map, as in \cref{fig:teaser}.



\noindent{\textbf{Lines}} were much less frequently discussed than the other mark types, with only four examples. All triples discuss the use of lines to show temporal data. 



\noindent{\textbf{Color}} is the most frequently discussed design concept. Answers in the dataset often make reference to specific color schemes or specific individual colors chosen, with rationales justifying these choices.
Rationales often make reference to data type. Several surfaced rationales also discussed semantic association of colors, for example using green to represent a positive category and red negative. Several students considered contrast and accessibility in their design choices.  Color was also discussed in reference to interaction, with triples discussing how colors change in response to user action, such as brushing. 
Only one triple in the dataset explicitly mentioned the decision not to use color encoding, emphasizing minimalism and visual clarity.

\noindent{\textbf{Position}}. Similarly to the points concept, questions in the dataset about position often probe which data attribute is being represented with position. The triples frequently mention encoding quantitative variables, such as in scatterplots. Some rationales reflected on how removing axis labels or scales could shift viewer focus from precise to relative comparisons. In several cases, the triples discuss geospatial position, again probing how geographical data was encoded.

\begin{figure}[ht!]
\includegraphics[width=1\columnwidth]{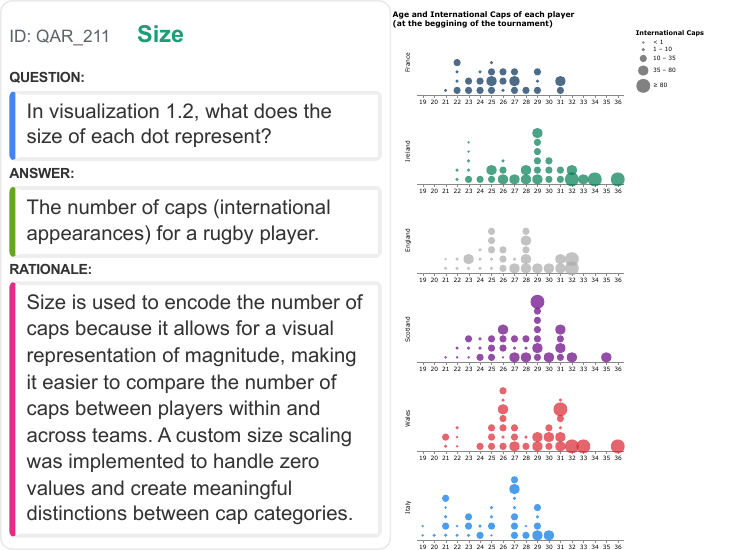}
  \caption{Example of a `Size' Question, Answer and Rationale from our dataset.}
  \label{fig:qar_size}
\end{figure}

\noindent{\textbf{Size}}. Triples about size often discuss using the size of points to encode quantitative data, as in \cref{fig:qar_size}. There is some overlap here with the area concept, as some students referred to this type of encoding as either area or size. As with area, the rationales often probed the use of nonlinear scaling to correct for perceptual issues.

\noindent{\textbf{Shape}} is infrequently discussed, with only 6 triples relating to this concept. Some triples discuss the use of shape to encode categorical variables. One rationale explicitly supported the decision not to use shape, noting that an additional encoding alongside both color and size could overwhelm the viewer.  

\noindent{\textbf{Interaction}} is commonly discussed by students. Questions often query the functionality of the interactive visualizations, such as tooltips, filtering, and zooming. The rationales frequently referenced Shneiderman's visual information seeking mantra \cite{shneiderman_eyes_2003}, discussing interactivity to support data exploration.

\noindent{\textbf{Layout}} is also commonly discussed, with questions generally probing how visualizations are spatially arranged. Rationales about faceting and juxtaposition were generally about supporting comparison. Some students described laying out visualizations sequentially, with rationales about narrative structure or storytelling. There were also several rationales about geospatial layouts, again, with discussions of relaxed geographies surfacing.

The distribution of QAR triples across design concepts is not uniform. This skew reflects the kinds of design rationales that students tend to foreground in their narratives.
Color is unsurprisingly well-represented as a versatile visual variable that can serve both functional and aesthetic roles across visualizations. 
Interaction and layout are also widely discussed, which is likely because they represent higher-level design concepts that transcend specific data types or visual idioms.Conversely, tilt is not represented in the dataset. This possibly reflects both the limited support for tilt-based encoding in Vega-Lite, and its general rarity in visualization design.  We note that its absence suggests how certain design concepts, though theoretically valid, may fall outside the practical or pedagogical scope of student-authored work.

\section{Discussion}


Our dataset provides a valuable resource for studying visualization design through natural language. It can support visualization recommendation systems by offering grounded examples of how specific design choices are justified in practice. It also has applications in design pedagogy, helping educators identify common reasoning patterns, gaps in understanding, or misconceptions among data visualization students. Furthermore, it can serve a benchmark for evaluating the visual understanding capabilities of multimodal LLMs from a design perspective.

We have carried out some preliminary experiments benchmarking Gemini 2.5 Flash \cite{noauthor_gemini_2025}, a state-of-the-art multimodal LLM. We prompted the model with the visualization(s) and question, and prompted  it to generate a free-form answer. The model's answers were evaluated against the correct answer using BERTScore F1 \cite{zhang_bertscore_2020},a metric popular in NLP which measures semantic similarity between generated and reference text on a scale of 0 to 1. Answers with a BERTScore of 0.85 or greater, indicating high semantic similarity, were considered correct. This threshold was confirmed through manual inspection. The model achieved $62\%$ accuracy rate, indicating that challenges remain for multimodal LLMs in accurately interpreting visualizations.

We also highlight that this dataset is not a neutral artifact. It reflects the pedagogical, personal and social contexts in which it was produced. Students developed their visualizations in response to course materials that, while grounded in theory, introduce concepts through a particular instructional lens. Students selected their own analysis datasets, resulting in subjects that reflect their individual lived contexts.
However, such contextual entanglement is not a limitation but a characteristic of the dataset. Design rationale does not arise in isolation — it is always entangled with, and thus shaped by the underlying data, available tools, and the designer's own context \cite{akbaba_entanglements_2025}. Future research should leverage this dataset as an artifact of real-world practice.

Nevertheless, the scope of the dataset does limit its generalizability. It does not capture how more experienced designers reason about visualization, nor how such reasoning might differ across tools or domains. However, the methodology we present for surfacing design rationale from narrative text is not tool- or population-specific. Future work could apply this approach to broader contexts to explore how design reasoning varies with experience and setting.


\newpage

\section*{Supplementary material}
The dataset is made available at \url{https://github.com/maevehutch/DesignQAR} underr a CC BY 4.0 license with an interactive viewer at \url{https://maevehutch.github.io/DesignQAR/}.

\section*{Acknowledgements}
We are grateful to Dr. Andrew Macfarlane, Chair of the Computer Science Research Ethics Committee at City St George’s, whose guidance was invaluable during the ethical approval process for this research.

\bibliographystyle{abbrv-doi}

\bibliography{template}
\end{document}